\documentclass[pdflatex,sn-nature]{sn-jnl}
\usepackage{graphicx}%
\usepackage{multirow}%
\usepackage{amsmath,amssymb,amsfonts}%
\usepackage{amsthm}%
\usepackage{mathrsfs}%
\usepackage[title]{appendix}%
\usepackage{xcolor}%
\usepackage{textcomp}%
\usepackage{manyfoot}%
\usepackage{booktabs}%
\usepackage{algorithm}%
\usepackage{algorithmicx}%
\usepackage{algpseudocode}%
\usepackage{listings}%
\usepackage{indentfirst}
\DeclareUnicodeCharacter{2212}{-}
\usepackage{abstract}
\renewcommand{\abstractname}{}

\title{Interorbital Antisymmetric Hopping Generated Flat Bands on Kagome and Pyrochlore Lattices}
\author[1]{\fnm{Keyu} \sur{Zeng}}\email{keyu.zeng@bc.edu}
\author*[1]{\fnm{Ziqiang} \sur{Wang}}\email{wangzi@bc.edu}
\affil[1]{\orgdiv{Department of Physics}, \orgname{Boston College}, \orgaddress{\city{Chestnut Hill}, \postcode{02467}, \state{Massachusetts}, \country{USA}}}

\begin{document}
\maketitle

\begin{abstract}
Flat bands are intriguing platforms for correlated and topological physics. Various methods have been developed to create flat bands utilizing lattice geometry, but the investigation of orbital symmetry in multiorbital materials is a new area of focus. Here, we introduce a site symmetry based approach to emerging multiorbital 2D and 3D flat bands on the kagome and pyrochlore lattices. As a conceptual advance, the one-orbital flat bands are shown to originate as mutual eigenstates of isolated molecular motifs. Further developing the mutual eigenstate method for multiple orbitals transforming differently under the site symmetries, we derive interorbital hopping generated flat bands from the antisymmetric interorbital Hamiltonian and introduce group-theoretic descriptions of the flat band wavefunctions. Realizations of multiorbital flat bands in realistic materials are shown to be possible in the Slater-Koster formalism. Our findings provide new directions for exploring flat band electronic structures for novel correlated and topological quantum states.
\end{abstract}

\section{Introduction}
    The search, discovery, and design of flat bands in electronic structure have attracted much interest for realizing novel correlated and topological states of matter due to the suppression of the kinetic energy. A primary example is the theoretical proposal of flat band enabled fractional quantum anomalous hall effect \cite{han_fractionalized_2012,sheng_fractional_2011,wang_fractional_2011,neupert_fractional_2011,regnault_fractional_2011}, which has been observed recently in twisted bilayer MoTe$_2$ \cite{park_observation_2023}, twisted bilayer graphene \cite{xu_observation_2023}, and multilayer graphene \cite{lu_fractional_2024}. There are diverse and intriguing physical phenomena associated with flat bands in quantum materials, including negative orbital flat band magnetism in kagome magnetic metal Co$_3$Sn$_2$S$_2$ \cite{yin_negative_2019}, strange metallicity in flat band kagome materials Ni$_3$In \cite{ye_hopping_2024}, and non-Fermi liquid behavior in flat band pyrochlore materials CuV$_2$S$_4$ \cite{huang_non-fermi_2024}. In the much attractive ``135'' family of kagome metals and superconductors $A$V$_3$Sb$_5$ ($A=$ K, Cs, Rb) and related compounds such as CsTi$_3$Bi$_5$, the interorbital flat band physics may play an important role as the low-energy electronic structure evades the one-orbital model \cite{hu_rich_2022, bandstructure} and incipient flat bands beyond density functional theory have even been detected recently \cite{yang_observation_2023, luo_van_2024}. Remarkably, the flat bands in the new member of the family CsCr$_3$Sb$_5$ are close to the Fermi level and produce a landscape of correlated magnetic and charge density waves states, and quantum criticality, non-Fermi liquid metals, and superconductivity under pressure \cite{liu_superconductivity_2024, xie_electron_2024}.
    Since the multiorbital nature is generic to quantum materials \cite{kim_signature_2022, bandstructure}, it is crucial to understand the multiorbital origin of the flat bands and their electronic wavefunctions, which are necessary for engineering flat bands and constructing theoretical models including correlations beyond band theory to study emergent novel electronic states.
    
    An increasing number of theoretical approaches have been developed to understand and construct flat bands \cite{leykam2018artificial,calugaru_general_2022,neves_crystal_2024}, including the line-graph method \cite{morfonios_flat_2021,chiu_fragile_2020,liu_orbital_2022,nakai_perfect_2022}, the Wannier function and the compact localized states method \cite{maimaiti_compact_2017,maimaiti_flat-band_2021,maimaiti_universal_2019, graf_designing_2021, chen_decoding_2022, rhim_singular_2021, hwang_general_2021}. Fine-tuning parameters in certain systems can also result in a flat band \cite{mizoguchi_flat-band_2019,ogata_methods_2021,lee_hidden_2019,misumi_new_2017}. 
    In this work, we present a systematic theory to construct single- and multi-orbital flat bands in the 2D kagome and 3D pyrochlore lattices made of corner sharing molecular motifs of triangles and tetrahedrons. Different and complementary to the existing approaches, we develop a mutual eigenstate method (MEM) based on the isolated molecular states \cite{mizoguchi_flat_2019, mizoguchi_flat-band_2019, mizoguchi_flat-band_2021, mizoguchi_systematic_2020}, which enables us to obtain the analytical flat band wavefunction and determine its
    group theoretic origin from the molecular states. Moreover, the MEM
    reveals the important correlation of the flat band wavefunction singularity and topological properties with the hopping symmetries within molecular motifs. We directly utilize the orbital symmetry at the lattice sites, i.e. the local site symmetry, which is particularly suitable for the construction of flat bands in the presence of multiple orbitals \cite{sun_nearly_2011,venderbos_narrowing_2011,mizoguchi_systematic_2020,calugaru_general_2022}. The new method is also applicable for treating equal number of orbitals on different sublattices, thereby contributing an important component to the recently proposed general construction of flat bands for unbalanced number of sublattice orbitals \cite{calugaru_general_2022}.
   
    We discover that the local site symmetry, such as mirror and inversion, plays a vital role. The mathematical origin of the interorbital hopping generated flat bands ("interorbital flat band" is used in the following text for brevity) is the formation of antisymmetric (or skew-symmetric) off-diagonal interorbital hopping Hamiltonian. We show that inversion even/odd orbitals give rise to singular flat bands containing band-touching points with dispersive bands, while mirror even/odd orbitals lead to a new type of nonsingular flat band in lattices with an odd-number of sublattice sites including the kagome lattice. The different singularity of the mirror/inversion flat bands arises due to whether the interorbital hopping breaks the symmetries within the molecular motifs. We demonstrate the realization and topological properties of the mirror-symmetry generated flat bands for multiorbital materials on the kagome lattice using the Slater-Koster formalism \cite{slater_simplified_1954}. 
    An intriguing property we discover is that the interorbital flat band can be transformed into a dispersive kagome band with inherited pure-type (p-type) van Hove singularity (vHS) where the wavefunctions are localized on one of the three sublattices in space. Beyond advancing the general theory of flat band construction, our findings are directly relevant for the intriguing electronic structure of the ``135'' kagome metals $A$V$_3$Sb$_5$ and related compounds, particularly the anomalous double p-type vHS \cite{hu_rich_2022, bandstructure} and the incipient flat bands near vHS detected in both CsTi$_3$Bi$_5$ \cite{yang_observation_2023} and CsV$_3$Sb$_5$ \cite{luo_van_2024}. Moreover, the interorbital flat band environment captures that in the pyrochlore CuV$_2$S$_4$ and provides a concrete model for including strong correlation effects beyond band theory \cite{huang_non-fermi_2024}.
    
    \begin{figure}[t]
	\centering
	\includegraphics[width=0.8\columnwidth]{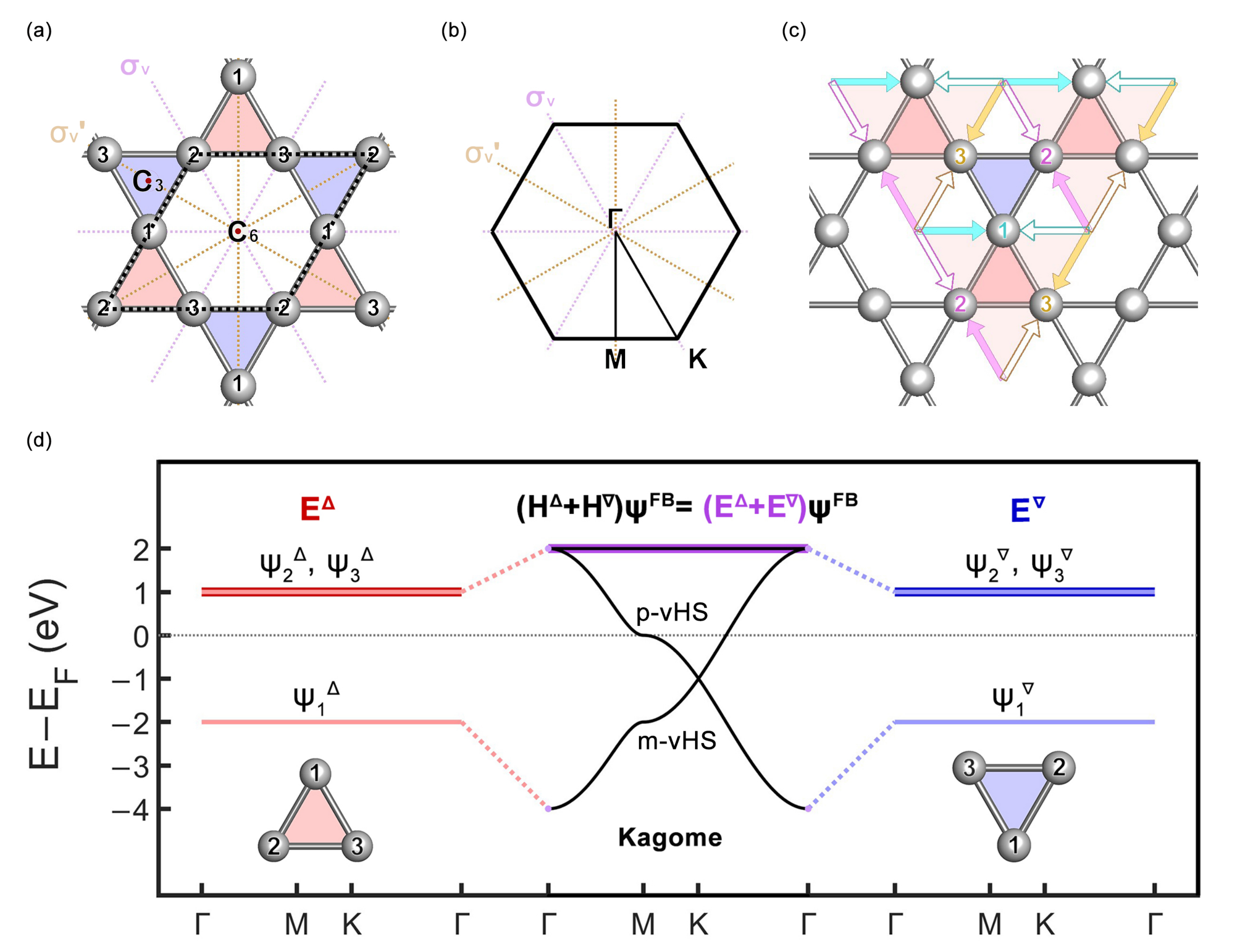}
	\caption{Flat band in one-orbital tight-binding model on the kagome lattice. (a) We use the $C_{6v}$ point group for the 2D kagome lattice and $C_{2v}$ group for the local site symmetry at Wyckoff position $3c$. There are two sets of mirror planes perpendicular to the kagome lattice. The dash lines indicate the unit-cell. (b) Brillouin Zone and high symmetry path of the kagome lattice. (c) Illustration of mutual eigenstate shared by up and down triangles for one orbital flat band. The arrows (filled: +; empty: -) indicate Fourier transformation vectors $\mathbf{r_{1/2/3}}-\mathbf{r^0}$ from three hexagonal centers. The sublattices $1/2/3$ are marked by cyan, magenta and yellow, respectively. (d) Illustration of kagome flat band construction (t=-1.0 eV) from isolated up (red) and down (blue) triangles with $\mathbf k$-independent energies.}
	\label{fig1}
    \end{figure}
    \section{One-orbital Model and MEM}
     The one-orbital tight-binding Hamiltonian with hopping $t$ on the kagome lattice can be written as $H=t\cdot H^K+\mu I$ in the 3-sublattice basis $\left[c^\dagger_1(\mathbf{k}), c^\dagger_2(\mathbf{k}), c^\dagger_3(\mathbf{k})\right]$, and
    \begin{equation}
    H^K = 2\left[\begin{array}{ccc}
    0 & \mathbf{c}(\mathbf{k}\cdot \mathbf{r}_{21}) &  \mathbf{c}(\mathbf{k}\cdot \mathbf{r}_{31})\\
    \mathbf{c}(\mathbf{k}\cdot \mathbf{r}_{12}) & 0 &  \mathbf{c}(\mathbf{k}\cdot \mathbf{r}_{32})\\
    \mathbf{c}(\mathbf{k}\cdot \mathbf{r}_{13}) & \mathbf{c}(\mathbf{k}\cdot \mathbf{r}_{23}) & 0 
    \end{array}\right]
    \label{Hk}
    \end{equation}
    where $\mu$ is the chemical potential, $\mathbf{r}_{ij}$ denotes the hopping vector connecting the sublattice sites $i$ and $j$ in an up triangle in Fig.~\ref{fig1}(a), and the wavevector $\mathbf{k}$ is defined in the first Brillouin zone in Fig.~\ref{fig1}(b). For brevity, 
    $\mathbf{c}(x) \equiv \cos(x)$ and $\mathbf{s} (x) \equiv \sin(x)$ are used throughout.
    It is widely known that the band structure obtained from the eigenstates of Eq.~(\ref{Hk}) contains a flat band, as shown in Fig.~\ref{fig1}(d, middle). Various methods, including the compact localized states based on Wannier functions \cite{hwang_general_2021} and line-graph of honeycomb lattice \cite{mielke_ferromagnetic_1991}, have been employed to understand the physical and mathematical origin of the flat band.
    
    Since the kagome lattice is made of alternating corner-sharing up and down triangles, the Hamiltonian $H^K$ in Eq.~(\ref{Hk}) can be decomposed as $H^K = H^\Delta + H^\nabla$,
    \begin{equation}
    H^{\Delta/\nabla}(\mathbf{k}) = \left[\begin{array}{ccc}
    0 & e^{\mp ik_3} &  e^{\pm ik_2}\\
    e^{\pm ik_3} & 0 &  e^{\mp ik_1}\\
    e^{\mp ik_2} & e^{\pm ik_1} & 0 
    \end{array}\right],
    H_{\text{molecule}}^{\Delta/\nabla} = \left[\begin{array}{ccc}
    0 & 1 & 1\\
    1 & 0 & 1\\
    1 & 1 & 0 
    \end{array}\right],
    \label{Hupdown}
    \end{equation}
    where $k_l =\epsilon_{ijl}\mathbf{k}\cdot \mathbf{r}_{ij}$. $H^\Delta=(H^\nabla)^*$ is a result of
    the inversion symmetry and the $H_{\text{molecule}}^{\Delta/\nabla}$ is the Hamiltonian for an isolated triangle molecule in the basis $\left[c^\dagger_1, c^\dagger_2, c^\dagger_3\right]$. We are thus motivated to develop a systematic molecular approach \cite{mizoguchi_flat_2019,mizoguchi_flat-band_2021} to the flat band states from the localized states on the isolated motifs. The eigenvalues of $H^{\Delta/\nabla}$ are independent of the reciprocal vector $\mathbf{k}$ with $n^2$ accidental degeneracy at each energy: $E_1 = 2$ in the $\mathbb{A}_1$ irreducible representation (irrep) and 
    $E_{2/3} = -1$ (irrep $\mathbb{E}$), corresponding to $n^2$ isolated triangle molecules. Combining the two non-commuting $H^\Delta$ and $ H^\nabla$, a flat band arises as a solution of the total Hamiltonian $H^K$ when there exists a mutual eigenstate $H^{\Delta/\nabla}\vert\Psi_{MEM}\rangle = E^{\Delta/\nabla}\vert\Psi_{MEM}\rangle$. The resulting $\mathbf{k}$-independent total energy $E_{FB} = E_{MEM}^\Delta+E_{MEM}^\nabla$ as indicated by the purple flat dispersion in Fig.~\ref{fig1}(d).

    The energy and wavefunction of the flat band can be directly derived from the $C_{3v}$ symmetry of the triangles combined with the $C_{2v}$ site symmetry subgroup of point group. The detailed analysis is given in Methods. For example, the wavefunction can be constructed using symmetry arguments and understood as a linear combination of the degenerate eigenstates in the $\mathbb{E}$ irrep $\Psi^\mathbb{E}_{2/3}$. The flat band solution requires the two-sublattice eigenvector $\vert\Psi_{3a}\rangle=\left[1,-1,0\right]/\sqrt{2}$ as part of the mutual eigenstate while the orthogonal eigenvector $\vert\Psi_2\rangle=\left[-1,-1,2\right]/\sqrt{6}$ becomes dispersive. The two degenerate states can be linearly combined into $\vert\Psi_{3b}\rangle=\left[-1,0,1\right]/\sqrt{2}$ and $\vert\Psi_{3c}\rangle=\left[0,1,-1\right]/\sqrt{2}$, forming a $C_3$ rotation symmetric two-sublattice eigenvector set. The $\vert\Psi_{3a/b/c}\rangle$ isolated molecular states are Fourier transformed with respect to three origins at the center of the adjacent hexagons as shown in Fig.~\ref{fig1}(c), and linearly combined in $1:1:1$ ratio to get the final wavefunction $\vert\Psi_{MEM}\rangle$. At $\Gamma$ point, $\vert\Psi_{MEM}\rangle$ vanishes and the degeneracy in $\mathbb{E}$ is recovered by $\vert\Psi_{2/3}\rangle$. The discontinuous jump of wavefunctions corresponds to the incomplete CLS as a criterion of the flat band singularity \cite{rhim_classification_2019, rhim_singular_2021, mizoguchi_flat_2019, mizoguchi_flat-band_2019}. For a general $\mathbf{k}$ point except $\Gamma$, the $\vert\Psi_{MEM}\rangle$ wavefunction is a basis function of the $\mathbb{B}_1$ irrep of the local site symmetry $C_{2v}$ with $n_\mathbf{k}^2-1$ accidental degeneracy, which originates from the $\mathbb{E}_1$ irrep in the induced representation 
    ${\rm Ind}(\mathbb{E})_{C_{6v}} = \mathbb{E}_1+\mathbb{E}_2$ from molecular hopping lattice point group $C_{3v}$ to the full kagome lattice point group $C_{6v}$. The topological nontrivial singularity of the flat band can also be understood by considering that the $\mathbb{B}_1$ irrep of the flat band cannot be induced to the $\mathbb{E}_2$ irrep of the band touching point. This concept of identifying singularities through symmetry is akin to the topological criterion used for disconnected elementary band representations (EBRs) \cite{bradlyn_topological_2017, cano_topology_2018, cano_building_2018}, but with the added consideration of a molecular symmetry group in addition to the site symmetry group. Notably, this approach avoids the need to calculate Wannier functions or identify EBRs for tight-binding models. 
    
    The detailed analysis provided in Methods leads to
        \begin{eqnarray}
        E_{FB}&=&E_{MEM}^\Delta+E_{MEM}^\nabla=-2 \\
        \vert\Psi^{FB}_K\rangle &=& \vert\Psi_{MEM}\rangle =\left[\mathbf{s}(k_1), \mathbf{s}(k_2), \mathbf{s}(k_3)\right]/N.
        \label{psi-mem}
    \end{eqnarray}
    for $H^K$, where $N$ is the normalization factor. The odd parity of $\vert\Psi^{FB}_K\rangle$ is a result of the alternating sign of the two-sublattice wavefunction $\vert\Psi_{3a/b/c}\rangle$. The single-branch flat band with a double-degeneracy touching point at $\Gamma$ in an one-orbital tight binding model for the kagome lattice is thus explained.
    
    \section{Inversion Interorbital Flat Band}
    We next extend the molecular MEM to construct interorbital flat bands for systems involving two orbitals. Properly orienting the orbitals based on kagome lattice symmetries, the diagonal blocks of the Hamiltonian for intraorbital hopping have the same form as $H^K$, with orbital-dependent hopping parameters $t_{11/22}$. The off-diagonal interorbital hopping blocks are dependent on the site symmetry of the orbitals. If the two orbitals belong to the same irrep in the site symmetry group, the interorbital hopping will be of the same form as $H^K$. In this case, absent of intraorbital hopping, the off-diagonal blocks generate two (a bonding and an anti-bonding) sets of typical kagome band dispersions with an effective interorbital hopping $t^{\pm} = \pm t_{12}$. The flat bands have identical properties with the one-orbital kagome flat band.
    
    New phenomena arise when the two orbitals $O_{1/2}$ transform differently (for instance, even or odd) under certain symmetry operations $S$ and thus belong to different irreps $\gamma_{1/2}$ of the site symmetry group: $\chi^{\gamma_{1/2}}(S) = \pm 1$ as an example. It is important to note that although the even and odd orbitals belong to different irreps of the local site symmetry, mixing between the orbitals are no longer forbidden at general $\mathbf{k}$-points in the 2D Brillouin zone because of the lowering of symmetry \cite{slater_simplified_1954}. We first consider the case where $S$ corresponds to the inversion operation. For instance, common inversion even/odd combination of orbitals can be $s$ and $p$ orbitals, or $d$ and $p$ orbitals. Inversion even and odd orbitals have been studied on the square lattice \cite{sun_nearly_2011} but with more hoppings beyond the nearest neighbor and parameter tuning. In lattices made of corner-sharing motifs, an exact flat band solution can be achieved. To satisfy the inversion symmetries, the off-diagonal inter-orbital hopping blocks of the two-orbital Hamiltonian must be antisymmetric (skew-symmetric): 
    \begin{equation}
    H^{oe}_{6\times6} = \left[\begin{array}{cc}
    t_{11}H_{11}^K & t_{12}H_{12}^{AS \dagger} \\
    t_{12}H_{12}^{AS} & t_{22}H_{22}^K
    \end{array}\right],
    \label{Hoe6by6}
    \end{equation}
    where
    \begin{equation}
    H^{AS}_{12}\equiv H^{AS}_I = 2i\left[\begin{array}{ccc}
    0 & -\mathbf{s}(k_3) &  \mathbf{s}(k_2)\\
    \mathbf{s}(k_3) & 0 &  -\mathbf{s}(k_1)\\
    -\mathbf{s}(k_2) & \mathbf{s}(k_1) & 0
    \end{array}\right].
    \end{equation}
    
    Mathematically, all odd dimension antisymmetric matrices must have at least one zero eigenvalue because their determinants are zero. That happens at all $\mathbf{k}$ gives rise to flat bands at zero-energy. Consequently, the second nearest neighbor interorbital hopping with the same sign structure also gives flat bands at $E = 0$. Similar zero-energy flat band was found in systems with spin-orbit-coupling \cite{ma_spin-orbit-induced_2020} interpreted by line-graph and split-graph, but not in multi-orbital systems.
    The antisymmetric interorbital $H_I^{AS}$ can also be divided into up and down triangles: $H^{AS}_I = H^\Delta_I + H^\nabla_I$. Because the triangles are related by inversion, $H^{AS}_I = H^\Delta - H^\nabla$ can be understood as flipping the sign of hopping $t_{12}$ for one set of triangles. Thus, the mutual eigenvector $\vert\Psi_{MEM}\rangle$ in Eq.~(\ref{psi-mem}) for the one-orbital Hamiltonian is also shared by the up and down triangles for $H^{AS}_I$, leading to two degenerate flat bands at energy
    \begin{equation}
    E_I^{FB} = E_{MEM}^\Delta-E_{MEM}^\nabla = 0,
    \end{equation}
    described by flat band wavefunctions
    \begin{align}
        \vert\Psi_{O_1}^{FB,I}\rangle & = \left[\mathbf{s}(k_1), \mathbf{s}(k_2), \mathbf{s}(k_3), 0, 0, 0\right]/N_0, \nonumber \\
        \vert\Psi_{O_2}^{FB,I}\rangle &= \left[0, 0, 0, \mathbf{s}(k_1), \mathbf{s}(k_2), \mathbf{s}(k_3)\right]/N_0.
        \label{i-FB}
    \end{align}
    
\begin{figure}[t]
	\centering
	\includegraphics[width=\columnwidth]{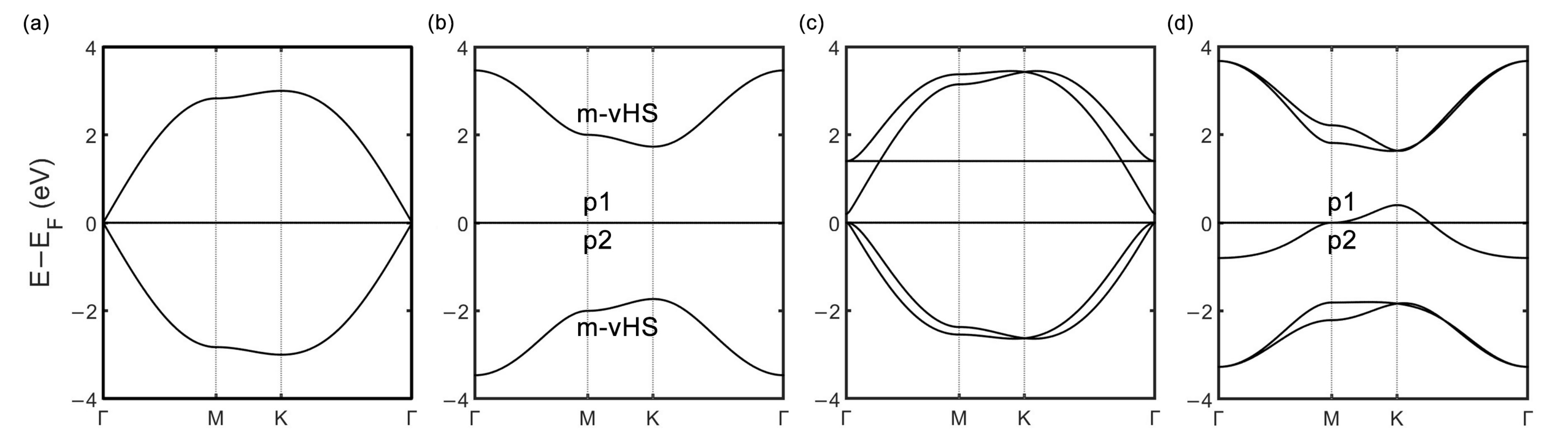}
	\caption{Band dispersions with inversion/mirror antisymmetric interorbital hopping generated flat bands on the kagome lattice. (a-b) Band dispersion of the two-orbital models showing singular inversion (a) and the nonsingular mirror (b) interorbital flat bands. The interorbital hopping is $t_{12}=1.0$ eV. Bands are doubly degenerate in the absence of intraorbital hopping. (c-d) Band dispersion in the inversion interorbital model, in the presence of nearest neighbor (nn) intraorbital hopping $t^{\rm nn}_{11}=-0.20$ eV (c) and the mirror interorbital model, in the presence of nearest neighbor intraorbital hopping $t^{\rm nn}_{22}=0.20$ eV (d).
    }
	\label{fig2}
    \end{figure}

    The flat bands with band-touching singularity are shown in the two-orbital band dispersion in Fig.~\ref{fig2}(a). Note that the odd-parity interorbital flat band wavefunctions in Eq.~(\ref{i-FB}) are also mutual eigenstates of the single-orbital flat band in each orbital sector given in Eq.~(\ref{psi-mem}). Consequently, two non-degenerate and perfectly flat bands remain robust even in the presence of significant intraorbital hopping, as shown in Fig.~\ref{fig2}(c). The singularity of the flat band at $\Gamma$ point evolves from Dirac-like under predominant interorbital hopping (Fig.~\ref{fig2}a) to quadratic band touching when intraorbital hopping becomes significant (Fig.~\ref{fig2}c). Moreover, different inversion symmetries usually involve orbitals of different angular momentum such as the $p$ and $d$ orbitals. The different atomic energies and crystal fields can further separate the singular flat bands of with these unique signatures in realistic materials of kagome as well as pyrochlore materials to be discussed below. 

    \section{Mirror Interorbital Flat Band}
    
    Next, we study the case where the site symmetry $S$ is with respect to all mirror operations and construct the interorbital flat bands. Consider  a mirror-even orbital and a mirror-odd orbital under $S$ with respect to either set of mirror planes perpendicular to the kagome lattice plane ($\sigma_v$ and $\sigma'_v$). Examples include the $p_x$ and $p_y$ orbitals or the $d_{xz}$ and $d_{yz}$ orbitals described by $\chi^{\mathbb{B}_1}_{p_x/d_{xz}}(\sigma_v/\sigma_{v'}) = \pm 1$ and $\chi^{\mathbb{B}_2}_{p_y/d_{yz}}(\sigma_v/\sigma_{v'}) = \mp 1$. The resulting interorbital hopping Hamiltonian matrix $H^{AS}_{12}\equiv H^{AS}_M$ must also be antisymmetric, but real and inversion-even: 
    \begin{equation}
    H^{AS}_M = 2\left[\begin{array}{ccc}
    0 & -\mathbf{c}(k_3) &  \mathbf{c}(k_2)\\
    \mathbf{c}(k_3) & 0 &  -\mathbf{c}(k_1)\\
    -\mathbf{c}(k_2) & \mathbf{c}(k_1) & 0
    \end{array}\right].
    \end{equation}
    The mirror antisymmetric interorbital $H^{AS}_M$ has emergent zero energy flat bands just as  the inversion-antisymmetric $H^{AS}_I$ discussed above. Thus, $E_M^{FB}=0$. There are, however, important differences in their properties. As shown in Fig.~\ref{fig2}(b), the mirror antisymmetric flat bands are no longer singular, i.e. without symmetry protected band-touching points with dispersive bands, in contrast to the inversion antisymmetric flat bands shown in Fig.~\ref{fig2}(a). Moreover, they are different from the flat bands constructed based on chiral operators \cite{calugaru_general_2022} since the interorbital flat bands in this work do not originate from site number differences but rather from the differences in orbital symmetry. 
    
    The flat band wavefunctions can be directly deduced from the lattice harmonics of $H^{AS}_M$. For a matrix of the form: $H^{AS}_{3\times3} = [0,a,-b;-a,0,c;b,-c,0]$, its zero-energy eigenvector wavefunction is $\vert\Psi^{AS}_{0}\rangle = \left[c,b,a\right]/N_0$, where $N_0$ is a normalization factor. Therefore, in the absence of intraorbital hopping,$\vert\Psi^{AS}_{0}\rangle$ is a zero-energy eigenvector of either orbital of the two-orbital Hamiltonian, leading to the flat band wavefuctions
    \begin{align}
        \vert\Psi_{O_1}^{FB,M}\rangle & = \left[\mathbf{c}(k_1), \mathbf{c}(k_2), \mathbf{c}(k_3), 0, 0, 0\right]/N_0, \nonumber \\
        \vert\Psi_{O_2}^{FB,M}\rangle &= \left[0, 0, 0, \mathbf{c}(k_1), \mathbf{c}(k_2), \mathbf{c}(k_3)\right]/N_0.
        \label{m-FB2}
    \end{align}
    They are parity even under mirror operation. 
    When intraorbital hoppings are added to one of the orbitals, the flatness and pure orbital content of the other orbital's flat band are unchanged, as shown in Figs.~\ref{fig2}(d).
    
    The interorbital flat band wavefunction can also be understood using the MEM. The two-orbital Hamiltonian can be divided into four chiral hopping sectors (clockwise: L; counterclockwise: R) on the up/down triangles: $H^{oe}_{M} = H_R^{\Delta}+H_L^{\Delta}+H_R^{\nabla}+H_L^{\nabla}$ (see Methods for details). For the mirror interorbital model, $t_R^{\Delta}=-t_L^{\Delta}=t_R^{\nabla}=-t_L^{\nabla}=1$. For each Hamiltonian, there are three degenerate bonding/anti-bonding eigenstates at energies $E_{\pm} = \pm 1$ described by two-sublattice eigenvectors. A shared eigenvector exists for a combination of mirror ($\sigma_v$) related chiral up and down triangles $H_A=H_{R}^{\Delta}+H_{L}^{\nabla}$ and $H_B=H_{L}^{\Delta}+H_{R}^{\nabla}$ with $E^{A/B}_{\pm} = \pm(t_{R/L}^{\Delta}+t_{L/R}^{\nabla})=0$ and the corresponding wavefunction $\vert\Psi_{O_1}^{A/B}\rangle= \vert e^{\pm k_1},e^{\pm k_2},e^{\pm k_3},0,0,0\rangle$ and $\vert\Psi_{O_2}^{A/B}\rangle=\vert0,0,0,e^{\mp k_1},e^{\mp k_2},e^{\mp k_3}\rangle$. It can be shown that $H_A\vert\Psi^B_{O_i}\rangle =(H_B\vert\Psi^A_{O_i}\rangle)^*$ is a pure imaginary vector, thus leading to two flat band solutions $\vert\Psi^{FB,M}_{O_i}\rangle=\vert\Psi^A_{O_i}\rangle+\vert\Psi^B_{O_i}\rangle$, as given in Eq.~(\ref{m-FB2}), since $(H_A+H_B)\vert\Psi^{FB,M}_{O_i}\rangle=\mu \vert \Psi^{FB,M}_{O_i}\rangle$ as the dispersive cross terms cancel out.
    
    It is interesting to note that the mirror interorbital flat band wavefunction $\vert \Psi_{O_i}^{FB,M}\rangle$ is an even-parity counterpart of the single-orbital kagome flat band wavefunction $\vert\Psi^{FB}_K\rangle$ in Eq.~(\ref{psi-mem}) and the inversion interorbital flat band wavefunction $\vert\Psi_{O_i}^{FB,I}\rangle$ in Eq.~(\ref{i-FB}). Conceptually, the even parity counterpart remains a flat band solution because the hopping sign structure within a triangle motif matches with the one-orbital flat band wavefunction. The difference in the parity of the flat band wavefunctions of the inversion and mirror even/odd Hamiltonians reveals an intriguing mechanism for generating two kinds of flat bands. For singular flat bands with band-touching points with other dispersive bands, such as the ones in the single-orbital and the inversion interorbital band structures shown in Fig.~\ref{fig1}(d, middle) and Fig.~\ref{fig2}(a), the uniform hopping within the motifs gives rise to mutual eigenstates. By contrast, hopping from orbital $O_i$ to orbital $O_j$ with alternating signs within each motif breaks the original mirror symmetry of the isolated triangles, thus removing the mirror-protected degeneracy; but if the signs match the one-orbital flat band solution, there can be a branch of nonsingular dispersion-canceling eigenstates without band-touching points. In the CLS language \cite{rhim_classification_2019, rhim_singular_2021}, the wavefunction of $\vert\Psi_{O_i}^{FB,M}\rangle$ is a complete CLS with no singularity. In the EBR language \cite{bradlyn_topological_2017, cano_topology_2018, cano_building_2018}, the band representations of the flat band $\vert\Psi_{O_i}^{FB,M}\rangle$ are connected because they are equal to the representations of the symmetrized orbital $O_i$ at any $\mathbf{k}$ point, thus no singularity exists. The MEM understanding of the flat band wavefunction provides a bottom-up construction of the CLS or Wannier function and a direct determination of the symmetry properties with insights into the singularity of the flat band.
    
    The inversion interorbital flat band can also be understood in a similar way by decomposing the multiorbital Hamiltonian $H^{oe}_I=H_A-H_B$. Because of the minus sign, in order to cancel out the dispersive cross term $H_A\vert\Psi_B\rangle+H_B\vert\Psi_A\rangle$, a flat band wavefunction of $\vert\Psi^I_{FB}\rangle=\vert\Psi_A\rangle-\vert\Psi_B\rangle$ is required, which is the inversion-odd solution (see Methods for details) given in Eq.~(\ref{memproof}). From another perspective, it does not break the point group symmetry of the isolated triangular molecules, but rather the inversion symmetry relating the up and down triangles. Therefore, the original $\mathbb{E}$ degeneracy still exists, leading to singular flat bands. Inversion even/odd orbitals have different orbital angular momentum quantum number $l$, while mirror even/odd orbitals can come from the same orbital with different magnetic quantum number $m_l$ or orientations. Thus, on general grounds, large interorbital hoppings between mirror even/odd orbitals are more likely to appear, of which the properties are discussed in the following.
    
\section{Properties and Realizations}
    
    The two degenerate interorbital flat bands can be shifted in energy by a difference in the orbital potential energy $\mu_1$ and $\mu_2$ without disturbing the flatness because they are pure in orbital content. The double-degeneracy of the dispersive bands will be lifted as well. An interesting feature of the mirror even/odd flat bands is the pure-type (p-type) site-localized wavefunctions at the M point shown in Figs.~\ref{fig2}(b) and (d). In each orbital sector, this corresponds to the sublattice polarized p-type state that resides at the van-Hove singularity (VHS) in a single-orbital kagome band structure (Fig.~\ref{fig1}d) with nn hopping 
    \cite{kiesel_sublattice_2012}. In both $H^K$ and $H^{AS}_{M,I}$, the p-type vHS is pinned at the energy of the corresponding orbital. While the interorbital second nn hopping does not perturb the flat bands, the intraorbital nn or second nn hopping in $H_{ii}^K$ transforms the corresponding flat band in Fig.~\ref{fig2}(b) into a kagome-like dispersion with unchanged p-type vHS wavefunction and energy shown in Fig.~\ref{fig2}(d). These remarkable properties of dynamical generation of interorbital flat bands and the transformation to vHS turns out to be crucial for understanding the electronic structure of the "135" kagome metals. At the $M$ points, the well-known $H^K$ band structure possesses a pair of p-vHS and m-vHS for the dispersive bands (Fig.~\ref{fig1}d), while the flat band hosts the mixed two-sublattice eigenstate. However, in the DFT band structure of ``135'' kagome metals, the pair of low-energy kagome-like bands are anchored by a pair of double p-type vHSs for the dispersive d-bands \cite{hu_rich_2022}, thus defying a single-orbital effective model description. The mirror flat band with p-type eigenstate at the $M$ point provides a mechanism by transforming the interorbital flat band to the kagome-like dispersion with the additional p-vHS, thus producing the anomalous double p-type van Hove singularities in the electronic structure \cite{bandstructure}. Moreover, this mechanism also provides a plausible origin for the flat band like spectral buildup around the p-vHS observed in ARPES measurements \cite{yang_observation_2023, luo_van_2024}.

    \begin{figure}[t]
	\centering
	\includegraphics[width=\columnwidth]{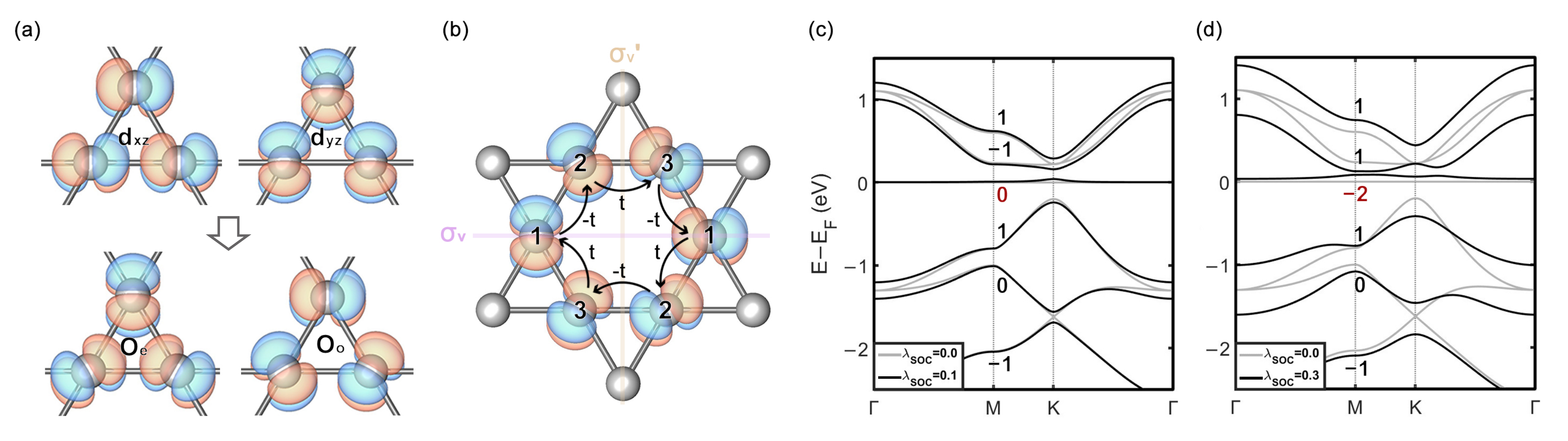}
	\caption{Realization of mirror interorbital flat band on the kagome lattice using the Slater-Koster formalism. (a) Linear combination of $d_{xz}$ and $d_{yz}$ orbitals to form two sets of orbital with different mirror symmetry for both sets of mirror planes. (b) Alternating two sets of orbitals on the kagome lattice have alternating signs of inter-orbital hopping parameters. $t_{\pi}=3t_{\delta}=0.6$ eV, $\mu_{1} = 0.0$ eV, $\mu_{2} = -1.0$ eV for (c) and (d). (c) Flat band disturbed by onsite spin-orbital-coupling with $\lambda_{SOC}=0.1$ eV, remaining topological trivial when there is no band-crossing. (d) Flat band disturbed by onsite spin-orbital-coupling with $\lambda_{SOC}=0.3$ eV, becoming topological non-trivial when band-crossing occurs.}
	\label{fig3}
    \end{figure}
    
    To make further connections to realistic materials \cite{riberolles_new_2024, ren_plethora_2022, guguchia_unconventional_2023}, we construct explicitly a two-orbital tight-binding Hamiltonian for the $d_{xz}$ and $d_{yz}$ orbitals using the Slater-Koster formalism \cite{slater_simplified_1954}. We show that although it is not enforced by the lattice symmetry and affected by intraorbital hopping, the interorbital flat band can play a significant role in the electronic structure of realistic materials. After linear combinations or rotations \cite{wu_nature_2021,liu_orbital_2022} of the two $d$-orbitals based on lattice symmetry, we obtain two hybrid orbitals labeled as 1 and 2, as shown in Fig.~\ref{fig3}(a), having different mirror symmetry (even and odd) with respect to the mirror planes $\sigma_v$ or $\sigma_{v'}$. The Hamiltonian $H_{xz/yz}$ has the structure of $H^{oe}_{6\times6}$ in Eq.~(\ref{Hoe6by6}), where the hopping parameters of the hybrid orbitals  determined by the overlap $t_\delta$ and $t_\pi$ for the $\delta$ and $\pi$ bonds: $t_{11} = (t_{\pi}-3t_{\delta})/2,$ $t_{22} = (t_{\delta}-3t_{\pi})/2$, and $t_{12} = \sqrt{3}(t_{\delta}+t_{\pi})/2$,
    \begin{equation}
    H_{xz/yz} = \left[\begin{array}{cc}
    \frac{1}{2}(t_{\pi}-3t_{\delta})H_{11}^K & \frac{\sqrt{3}}{2}(t_{\delta}+t_{\pi})H_{M}^{AS \dagger} \\
    \frac{\sqrt{3}}{2}(t_{\delta}+t_{\pi})H_{M}^{AS} & \frac{1}{2}(t_{\delta}-3t_{\pi})H_{22}^K 
    \end{array}\right].
    \label{H-SK}
    \end{equation}
    As illustrated in Fig.~\ref{fig3}(b), the interorbital hopping indeed has an alternating sign structure. Intriguingly, when $t_{\pi}=3t_{\delta}$ or $t_{\delta}=3t_{\pi}$, the intraorbital hopping for one of the hybrid orbitals in the diagonal blocks of Eq.~(\ref{H-SK}) vanishes. Our findings then imply that the associated interorbital flat band would remain perfect flat, while the other becomes significantly dispersive. This is confirmed by the calculated band dispersion plotted in gray lines in Fig.~\ref{fig3}(c) and (d), in the presence of a moderate crystal field splitting between the hybrid orbitals ``1'' and ``2''. We thus predict interorbital nearly flat or narrow bands involving $d_{xz}$ and $d_{yz}$ orbitals in proximity to having $t_\delta :t_\pi=1:3$ or $3:1$ in kagome materials.
    
    It is constructive to study the effects of atomic spin-orbit coupling (SOC) $H_{soc}=\lambda_{soc}\mathbf{L}\cdot\mathbf{S}$. Since the hybrid orbitals are $l=2$ angular momentum eigenstates with magnetic quantum number $m_l=\pm1$, the SOC leaves the spin component $s_z$ conserved and the up and down spin bands degenerate. For small $\lambda_{soc}$, the bands shifts and the band crossings are split as shown in Fig.~\ref{fig3}(c). 
    The flat band remains isolated and mostly across the whole Brillouin zone. The calculated Chern number for a single spin projection is marked next to each band, which is the same as the spin Chern number when taking into account the degenerate band of the different spin-projection carrying an opposite Chern number. In this case, the isolated flat band carries zero Chern number as proved in previous work on nonsingular flat bands \cite{chen_impossibility_2014, rhim_classification_2019}.
    Interestingly, increasing $\lambda_{soc}$ causes gap between the flat band and the dispersive band to close and reopen as plotted in Fig.~\ref{fig3}(d), and endows the flat band with a nontrivial spin-resolved Chern number or a spin Chern number.

    Topologically nontrivial flat bands have been investigated for possible realizations of fractionalized anomalous quantum states. When time-reversal symmetry is broken either spontaneously by correlation effects or by coupling to ferromagnetic structures, partial occupation of the a spin-polarized flat band has the potential for realizing fractional quantum anomalous Hall state or fractional Chern insulators \cite{kang_topological_2020,okamoto_topological_2022,han_fractionalized_2012,sheng_fractional_2011,wang_fractional_2011,neupert_fractional_2011,regnault_fractional_2011}. When time-reversal symmetry is preserved, partial filling of the degenerate spin-Chern band has the potential of realizing fractional quantum spin Hall state or the proposed fractional topological insulator \cite{sun_nearly_2011}. The isolated topological interorbital flat band discussed here can provide a useful direction for material realizations of the fractionalized quantum states. 
    
\section{Extension to Pyrochlore}
\begin{figure}[t]
	\centering
	\includegraphics[width=\columnwidth]{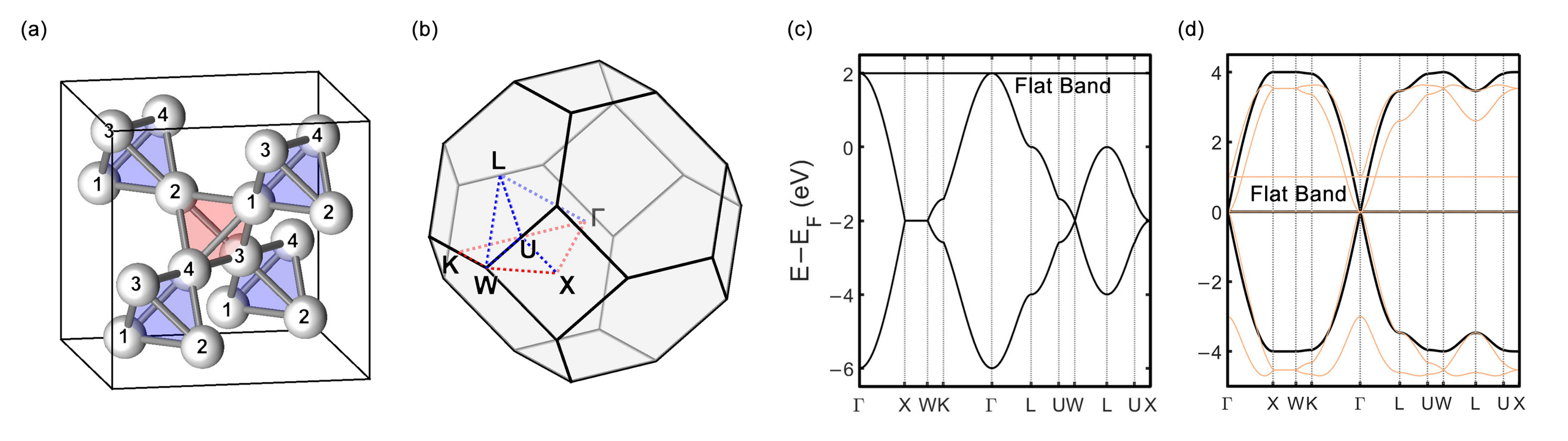}
	\caption{One-orbital and inversion interorbital flat bands on the pyrochlore lattice. (a) Pyrochlore lattice structure consisting of up and down tetrahedrons. The solid lines indicate the unit-cell. (b) Brillouin Zone and high symmetry path. (c) Pyrochlore one-orbital band structure with $t=-1.0$ eV, where there are doubly-degenerate flat bands. (d) Inversion interorbital flat band with $t_{12}=1.0$ eV and $t_1=0.0$ eV, $t_2=0.0$ eV (black thick line) or $t_1=-0.5$ eV, $t_2=0.0$ eV (orange thin line).}
	\label{fig4}
    \end{figure}

     The 3D pyrochlore lattices are made of isolated motif of apex-sharing tetrahedrons with 4-sublattices, as shown in Fig.~\ref{fig4}(a). The band dispersions of a single-orbital on the pyrochlore lattice with nearest neighbor hopping are plotted in Fig.~\ref{fig4}(c) along the high symmetry directions of the 3D Brillouin zone in Fig.~\ref{fig4}(b). Similar to the 2D kagome lattice, the 3D flat band wavefunction in the single-orbital pyrochlore can be understood as the mutual eigenstates of the up (red) and down (blue) tetrahedrons (see Methods Fig.~\ref{figS2}). Applying the MEM to the $\mathbb{T}_2$ irrep of the tetrahedron point group $T_d$, there are two sets of two-sublattice eigenvectors $\Psi^{P}_{j=1,2}$ that form a shared 3D flat band wavefunction at energy
    $E_{FB}^{P}=-2t$,
    \begin{align}
        \vert\Psi_{a}^{FB}\rangle & = \left[\mathbf{s}(k_{23}), \mathbf{s}(k_{31}), \mathbf{s}(k_{12}), 0\right]/N_a, \nonumber \\
        \vert\Psi_{b}^{FB}\rangle &= \left[\mathbf{s}(k_{24}), \mathbf{s}(k_{41}), 0, \mathbf{s}(k_{12})\right]/N_b
        \label{P-FB}
    \end{align}
    where $k_{ij}=\mathbf{k}\cdot (\mathbf{r}_i-\mathbf{r}_j)$. These parity-odd wavefunctions, responsible for the double-degeneracy of the 3D flat bands, are basis functions of the $\mathbb{E}_u$ irrep of the $D_{3d}$ site symmetry. The 3D flat bands have been observed experimentally \cite{wakefield_three-dimensional_2023}.

    A natural question is whether 3D interorbital flat bands can be constructed for multiorbital quantum materials on the pyrochlore lattice. Building on the findings on the kagome lattice, we consider two atomic orbitals that even and odd under a site symmetry operation $S$ on the pyrochlore lattice such as $s-p$, $p-d$ or $d-f$ combinations. Because of the four sublattices, the tight-binding Hamiltonian is now an $8\times8$ matrix of the same form as in Eq.~(\ref{Hoe6by6}). A crucially important difference from the kagome lattice is that the $4\times4$ off-diagonal antisymmetric interorbital hopping matrix ($H_{12}^{AS}$) is now even-dimensional. As a result, only inversion even/odd $H^{AS}_I$ is allowed, but mirror even/odd $H^{AS}_M$ vanishes in a two-orbital model due to the impossible sign alternations to satisfy all mirror or $C_2$ operations. The antisymmetric inversion even/odd interorbital Hamiltonian $H_{12}^{AS}\equiv H_I^{AS}$ on the pyrochlore lattice is given by
    \begin{equation}
    H_{12} = 2i \left[\begin{array}{cccc}
    0 & -\mathbf{s}(k_{21}) &  -\mathbf{s}(k_{31})  &  -\mathbf{s}(k_{41})\\
    \mathbf{s}(k_{12}) & 0 &  -\mathbf{s}(k_{32})  &  -\mathbf{s}(k_{42})\\
    \mathbf{s}(k_{13}) & \mathbf{s}(k_{23}) & 0  &  -\mathbf{s}(k_{43}) \\
    \mathbf{s}(k_{14}) & \mathbf{s}(k_{24}) & \mathbf{s}(k_{34})  &  0 \\
    \end{array}\right].
    \label{I-Pyrochlore}
    \end{equation}
    Note that the determinant of an even-dimensional antisymmetric matrix is a Pfaffian, which is usually non-zero. Surprisingly, the inversion interorbital matrix in Eq.~(\ref{I-Pyrochlore}) has a zero determinant $ {\rm Det}(H_{12}^{AS}) = 0$ due to the lattice geometry. 
    The flat band wavefunctions are give by
    \begin{align}
        \vert\Psi_{O_1,a}^{FB}\rangle & = \left[\mathbf{s}(k_{23}), \mathbf{s}(k_{31}), \mathbf{s}(k_{12}), 0, 0, 0, 0, 0\right]/N_a, \nonumber \\
        \vert\Psi_{O_1,b}^{FB}\rangle &= \left[\mathbf{s}(k_{24}), \mathbf{s}(k_{41}), 0, \mathbf{s}(k_{12}), 0, 0, 0, 0\right]/N_b,\\
        \vert\Psi_{O_2,a}^{FB}\rangle & = \left[0, 0, 0, 0, \mathbf{s}(k_{23}), \mathbf{s}(k_{31}), \mathbf{s}(k_{12}), 0\right]/N_a, \nonumber \\
        \vert\Psi_{O_2,b}^{FB}\rangle &= \left[0, 0, 0, 0, \mathbf{s}(k_{24}), \mathbf{s}(k_{41}), 0, \mathbf{s}(k_{12})\right]/N_b.
        \label{P-FB2}
    \end{align}
    In the inversion interorbital model, there is a new Dirac crossing at $\Gamma$ point in both kagome and pyrochlore lattices as shown in Fig.~\ref{fig2}(a) and Fig.~\ref{fig4}(d). Similar to the kagome lattice, the inversion interorbital flat bands on the pyrochlore lattice have the remarkable property that their flatness is robust against significant intraorbital hopping, which only shift the flat band energies, as explicitly shown in Fig.~\ref{fig4}(d). The same singularity and triple degeneracy at $\Gamma$ point can also be explained by the intact symmetry of the tetrahedrons, independent of the inversion symmetry that relates the up and down tetrahedrons.
    
    In realistic band structures, a hybrid of quadratic and Dirac crossing may occur. Whether the interorbital hopping is dominant or not can be judged by the features of the dispersive bands with respect to the singular flat band. For instance, in a recent study \cite{huang_non-fermi_2024} of CuV$_2$S$_4$, which contains a pyrochlore structure, the interorbital Dirac crossing features of the dispersive bands dominate over quadratic band touching at $\Gamma$ point, pointing to the proximity to an inversion interorbital flat band between different atoms (possibly of Cu $d$ and S $p$ orbitals) with significant interorbital hopping.

\section{Discussion}
    
    We introduced a new theoretical framework to discover and construct singular and nonsingular flat bands in multiorbital 2D kagome and 3D pyrochlore crystals. These lattice structures have corner-sharing motifs containing either an odd or an even number of sublattices, which are shown to be suitable for the isolated molecular approach and the mutual eigenstate method. The method bridges the gap between geometric understanding of the lattice based on line-graph methods and the numerical construction of orthonormal Wannier functions in realistic multiorbital electronic structures. In comparison to the compact localized state (CLS) method which interprets the flat band by constructing CLS from the eigenmode \cite{rhim_singular_2021}, our method directly derives the flat band eigenstates without diagonalizing the tight-binding Hamiltonian. Additionally, it establishes a connection between the symmetry of the flat band wavefunction and the underlying lattice geometry. For the molecular state method itself, we have advanced it by using the Mutual Eigenstate Method (MEM) to find the wavefunction, replacing the original mathematical construction \cite{mizoguchi_flat_2019, mizoguchi_flat-band_2019, mizoguchi_flat-band_2021, mizoguchi_systematic_2020}. This modification provides deeper insights into the degeneracy of flat bands, especially in higher dimensions such as the pyrochlore lattice. Although the current methods are limited to a small number of orbitals, they can serve as an elementary starting point for constructing CLS or Wannier states.
    
    We discovered that the local site-symmetry $S$ plays a crucial role for the emergence of interorbital flat bands between a pair of even/odd orbitals with respect to $S$. Interorbital flat bands and wave functions are found for $S$ corresponding to the local inversion and mirror symmetries on the kagome lattice, and for local inversion on the 3D pyrochlore lattice. The singularity properties and the energies of the flat bands are shown to be determined by the hopping symmetries within isolated molecular motifs. The singularity of flat bands can be classified by the incomplete CLS \cite{rhim_classification_2019, rhim_singular_2021} or similarly the absence of symmetry-preserving, exponentially localized Wannier functions \cite{bradlyn_topological_2017, cano_topology_2018, cano_building_2018}. Our approach offers both intuitive and analytical insights into flat band singularities by examining the symmetry properties of interorbital hopping within isolated molecular eigenstates. If the interorbital hopping preserves isolated molecular symmetries, the original degeneracy of molecular eigenstates ensures a singularity or band-touching of flat bands with the dispersive band at the $\Gamma$ point, corresponding to the incompleteness of CLS at the touching point. On the other hand, when the interorbital hopping breaks any isolated molecular symmetry, it lifts the degeneracy of the molecular eigenstate, resulting in nonsingular flat bands. Our method shares some similarities with the concept of identifying Elementary Band Representations (EBRs) for topologically nontrivial bands \cite{bradlyn_topological_2017, cano_topology_2018, cano_building_2018}, but it provides a compact, lattice geometry and orbital symmetry-based approach for multiorbital tight-binding models without calculating Wannier functions or the full determination of EBRs.
    
    Direct atomic realizations of such interorbital flat bands and the potential for hosting novel topological states are illustrated in the Slater-Koster framework for the mirror even/odd orbitals on the kagome lattice. The proposed mechanism for the interorbital flat bands is directly relevant to the search and design of flat bands in kagome and pyrochlore materials for studying novel correlated and topological quantum states. It provides a plausible interorbital flat band mechanism for the intriguing electronic structure of kagome metals $A$V$_3$Sb$_5$ such as the anomalous p-type vHS \cite{hu_rich_2022, bandstructure} and the flat band high spectral intensity buildup around the vHS below the Fermi level observed in CsTi$_3$Bi$_5$ \cite{yang_observation_2023} and CsV$_3$Sb$_5$ \cite{luo_van_2024}, as well as the incipient flat band responsible for the non-Fermi behavior in pyrochlore metal CuV$_2$S$_4$ \cite{huang_non-fermi_2024}. The findings pave the way for new directions in flat band exploration and the understanding of multiorbital electronic structures.

    \section{Data and Code Availability}
    Relevant codes and data generated in this work are available from the authors upon request.

    \section{Acknowledgements}

    We thank Jie Liu and Zhan Wang for valuable discussions. The work is supported by the U.S. Department of Energy, Basic Energy Sciences Grant DE-FG02-99ER45747 and by Research Corporation for Science Advancement Cottrell SEED Award No. 27856.

    \section{Author Contributions}
    Both K.Z. and Z.W. initiated the project, carried out the research and wrote the manuscript. All authors have read and approved the manuscript.

    \section{Competing Interests}
    The authors declare no competing interests.
    
\bibliography{citation}

\renewcommand{\thesection}{M}

\counterwithin{equation}{section}
\section{Methods}
\subsection{General Proof of the Symmetry Structure}
    The mirror symmetry operator $M$ along the y-direction that keeps site-1 invariant while exchanging site-2 with site-3 can be expressed as a $6\times6$ matrix in the two-orbital basis $C$ for the even/odd orbitals: 
    \begin{equation}
    M = \left[\begin{array}{cccccc}
    1 & 0 & 0 & 0 & 0 & 0 \\
    0 & 0 & 1 & 0 & 0 & 0 \\
    0 & 1 & 0 & 0 & 0 & 0 \\
    0 & 0 & 0 & -1 & 0 & 0\\
    0 & 0 & 0 & 0 & 0 & -1\\
    0 & 0 & 0 & 0 & -1 & 0
    \end{array}\right], \quad C = \left[\begin{array}{c}
    c^e_1\\
    c^e_2\\
    c^e_3\\
    c^o_1\\
    c^o_2\\
    c^o_3
    \end{array}\right]
    \label{Moperator}
    \end{equation}
     The other set of mirror symmetries can also be represented in similar ways. The Hamiltonian satisfies: $MHM^{-1} = H$. Therefore, the interorbital hopping has the antisymmetric form: $\epsilon_{ijk}t_{ij}c^e_ic^o_j$, where $\epsilon_{ijk}$ is the total antisymmetric tensor.
     
     Another way to describe the symmetry structure is based on the two-center integral approximation \cite{slater_simplified_1954}, where the hopping strength is approximated by the atomic orbital overlap $f^{\alpha\beta} = \int \Psi^\alpha\Psi^\beta dr^3$. The representation of the hopping parameters can be determined: $\chi^{\gamma}_{f^{\alpha\beta}}(S) = \chi^{\gamma_\alpha}_{\Psi^\alpha}(S)\chi^{\gamma_\beta}_{\Psi^\beta}(S)$, where $S$ is a local site symmetry.  For the kagome lattice, $\chi^{\gamma}_{f^{\alpha\beta}}(S)$ are all one-dimensional irreps and the intraorbital hopping for any orbital gives $\chi^{\gamma}_{f^{\alpha\beta}}(S) = 1$ and $\gamma = {\mathbb A}_1$. Therefore, all intraorbital hopping matrix exhibits simple s-orbital symmetry. For the interorbital hopping between, e.g. mirror even and odd orbitals, $\chi^{\gamma}_f = \chi^{{\mathbb B}_1}_{d_{xz}}\chi^{{\mathbb B}_2}_{d_{yz}}$, $\gamma = {\mathbb B}_1\otimes {\mathbb B}_2={\mathbb A}_2$. Thus, the interorbital hopping terms $t^{{\alpha}{\beta}}_{ij}c^{\alpha}_ic^{\beta}_j$ should transform under this irrep. For the pyrochlore lattice, the two-dimensional irreps become complicated with complex representations. For example, $\chi^{\gamma}_f = \chi^{\mathbb{E}_g}_{d_{xz}}\chi^{\mathbb{E}_g}_{d_{yz}}, \gamma = {\mathbb{E}}_g\otimes {\mathbb{E}}_g={\mathbb {A}}_{1g}+{\mathbb{A}}_{2g}+{\mathbb{E}}_{g}$.
     
    \subsection{MEM for Single Orbital Model}
     The eigenvectors of a triangle molecule are: $\vert\Psi_1\rangle=\left[1,1,1\right]/\sqrt{3}$, $\vert\Psi_2\rangle=\left[-1,-1,2\right]/\sqrt{6}$ and $\vert\Psi_{3a}\rangle=\left[-1,1,0\right]/\sqrt{2}$. The latter two degenerate states can be linearly combined into $\vert\Psi_{3b}\rangle=\left[1,0,-1\right]/\sqrt{2}$ and $\vert\Psi_{3c}\rangle=\left[0,-1,1\right]/\sqrt{2}$, which together with $\vert\Psi_{3a}\rangle$ form a $C_3$ rotation symmetric eigenvector set. Placing up (down) triangle molecules on a triangular net with aligned equal triangle edges $\mathbf{a}_i$ and shared vertexes results in the kagome lattice. The eigenvectors of $n^2$ molecular $H^\Delta$ and $H^\nabla$ in the kagome lattice is equivalent to a Fourier transform of the molecular eigenvectors based on the translation symmetry of the triangular lattice, while the eigenvalues remain degenerate and independent of $\mathbf{k}$:
     \begin{eqnarray}
     \vert\Psi^{\Delta/\nabla}_1\rangle&=&\left[e^{\pm ik_1}, e^{\pm i(k_1-k_3)}, e^{\pm ik_3}\right]/N_1, \\
     \vert\Psi^{\Delta/\nabla}_2\rangle&=&\left[-e^{\pm ik_1}, -e^{\pm i(k_1-k_3)}, 2e^{\pm ik_3}\right]/N_2, \\
     \vert\Psi^{\Delta/\nabla}_{3a} \rangle&=&\left[-e^{\pm ik_1}, e^{\pm ik_2}, 0\right]/N_3. 
     \end{eqnarray}
     The two sets of eigenvectors form the eigenspaces of the Hamiltonians of the up and down triangles, spanning the same $\mathbf{k}$-space, 
    \begin{equation}
    H^{\Delta/\nabla} = t^{\Delta/\nabla}\left[\begin{array}{ccc}
    0 & e^{\mp ik_3} &  e^{\pm ik_2}\\
    e^{\pm ik_3} & 0 &  e^{\mp ik_1}\\
    e^{\mp ik_2} & e^{\pm ik_1} & 0 
    \end{array}\right].
    \label{updownH}
    \end{equation}
    The Hamiltonians and eigenvectors of the up and down triangles are complex conjugate of each other due to the inversion symmetry relating the two sets of triangles. The degeneracy of the energies at $E = -t$ corresponds to the band-touching degenerate point at the $\Gamma$ point. Away from the $\Gamma$ point, the degeneracy is lifted and only one eigenvector still has dispersionless energy, which is the mutual eigenstate of $H^{\Delta/\nabla}$. To find the shared eigenspace, a linear combination of $\mathbf{k}$-space eigenvectors with different origins is performed, 
    \begin{align}
       \Phi^{\gamma_j}\mathbf{k}&=\sum^3_{m=1}\sum^{n_o}_{n=1}\sum^{d_j}_{i=1}e^{i\mathbf{k}\cdot(\mathbf{r_m}-\mathbf{r^0_n})}a_iw^{\gamma_j}_{i,m}c^\dagger_{k,m}\nonumber\\
       &=\sum^{n_o}_{n=1}\sum^{d_j}_{i=1}a_i\Psi^{\gamma_j}_{i}(\mathbf{k},n)
    \label{lceigv} 
    \end{align}    
    which is allowed by the energy degeneracy for all $\mathbf{k}$-point.
    Here $a_i$ is the linear combination coefficients between different eigenstates, $w^{\gamma_j}_{i,m}$ is the eigenstate coefficients for m-th site in the i-th eigenvector of irrep $\gamma_j$, and 
    $\{\mathbf{r^0_n}\}$ is a limited set of $n_o\ll n^2$ origins. 
    
    \begin{figure}[t]
	\centering
	\includegraphics[width=\columnwidth]{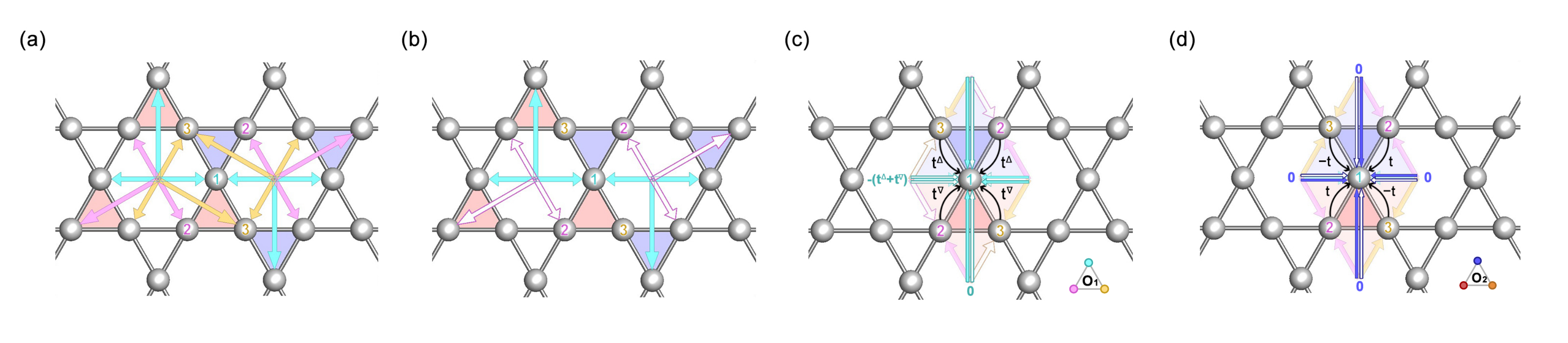}
	\caption{Demonstration of mutual eigenstate method (MEM) on the kagome lattice. Up/down triangles are marked in red/blue faces. $\vert\mathbf{r_{1/2/3}}-\mathbf{r^0_{a/b/c}}\vert$ belonging to different sublattices 1/2/3 are colored in cyan, magenta and yellow, and $w=\pm1$ is in filled and empty arrows, respectively. The curved black arrows indicate the hopping terms with hopping parameters $t^{\nabla/\Delta}$. (a) Schematics for $\vert\Psi^{\Delta}_1\rangle$ (left) and $\vert\Psi^{\nabla}_1\rangle$ (right). (b) Schematics for $\vert\Psi^{\Delta}_{3a/b/c}\rangle$ (left) and $\vert\Psi^{\nabla}_{3a/b/c}\rangle$ (right) before linear combination. (c) Symmetric hopping within triangles for one-orbital and inversion interorbital flat bands from site 2 or 3 to site 1. The schematic shows that the flat band eigenstate has a $\mathbf{k}$-independent eigenvalue $-(t^{\nabla}+t^\Delta)$. (d) Antisymmetric hopping for $\vert\Psi^{FB,M}_{O_1}\rangle$ from orbital $O_1$ (light colors) to site 1 orbital $O_2$ (dark colors).}
	\label{fig5}
    \end{figure}

    The mutual eigenvector satisfies invariance under inversion: $g_I\Phi^{\gamma_j}_{k}=\pm\Phi^{\gamma_j}_{g_Ik}$, each site remains themselves under inversion thus a set of inversion related $g_I(\mathbf{r_m}-\mathbf{r^0_n})=\mathbf{r_m}-\mathbf{r^0_{n'}}$ becomes the main requirement as shown in Fig~\ref{fig5}(c). By contrast, before linear combination, the $\vert\Psi^{\Delta}_{1/2/3}\rangle$ does not match with the right eigenvector $\vert\Psi^{\nabla}_{1/2/3}\rangle$ pattern because they are not inversion invariant as shown in Fig~\ref{fig5}(a) and (b). The problem is thus transformed into finding a proper set of origins and eigenvectors.
   
     For an origin $\mathbf{r^0_n}$, if there exists $\vert\mathbf{r_p}-\mathbf{r^0_n}\vert\neq\vert\mathbf{r_q}-\mathbf{r^0_n}\vert$ for non-zero $w^{\gamma_j}_{i,p/q}$ (number of $w$ can be two or three), there must exist another origin $\mathbf{r^0_{n'}}$ in order to form the inversion pair for site $p$ or $q$. Because there is no possibility for a one-sublattice eigenstate in $\mathbb{E}$, if $\vert\mathbf{r_p}-\mathbf{r^0_n}\vert=\vert\mathbf{r_p}-\mathbf{r^0_{n'}}\vert$, there must exists $\vert\mathbf{r_s}-\mathbf{r^0_{n'}}\vert\neq\vert\mathbf{r_p}-\mathbf{r^0_{n'}}\vert\neq\vert\mathbf{r_s}-\mathbf{r^0_n}\vert$. As a consequence, the number of origins quickly increase $n_o\rightarrow Z*N$, where $Z\leq D(C_{6v})$ and won't enclose. In a word, $\{\mathbf{r_m}-\mathbf{r^0_n}\} (m = {\vert w^{\gamma_j}_{i,m}\vert \neq 0})$ is not inversion invariant unless for any non-zero $w^{\gamma_j}_{i,p/q}, \vert\mathbf{r_p}-\mathbf{r^0_n}\vert=\vert\mathbf{r_q}-\mathbf{r^0_n}\vert$. The center of the triangles can also be excluded from the origin choice due to the infinite propagation of the inversion-pair vectors. As a result, only the eigenvectors $\vert\Psi_{3a/b/c}\rangle$ with two non-zero $w^{\gamma_j}_{i,p/q}$ (two-sublattice for short) satisfy the above requirement. Meanwhile the origins and their inversion partners have to lie on the three mirror planes $\sigma_v$ bisecting the bonds to satisfy the conditions simultaneously, which are the center of the hexagons. Around one triangle, there are three hexagon centers, forming a set of $\{\vert\mathbf{r_m}-\mathbf{r^0_n}\vert\}$ that contains all point group symmetry partners $g_{I}\vert\mathbf{r_m}-\mathbf{r^0_n}\vert$.
     
     Thus we have derived the flat band solution $\vert\Psi^{FB}_K\rangle$. Because the eigenvectors $\vert\Psi_{3a/b/c}\rangle$ are odd with respect to the mirror planes bisecting corresponding bonds, the combined eigenvector $\vert\Psi^{FB}_K\rangle$ is odd with respect to inversion symmetry as shown in Fig.~\ref{fig5}(c): $\vert\Psi^{FB}_K\rangle = \left[\mathbf{s}(k_1), \mathbf{s}(k_2), \mathbf{s}(k_3)\right]/N$. The description of ${\mathbb B}_1$ irrep is for the wavefunction under the site symmetry $C_{2v}$. The resulting wavefunction is the shared eigenstate of $H^\Delta$ and $ H^\nabla$ and has a $k$-independent total energy $E^{FB}_K = E_{3}^\Delta+E_{3}^\nabla = -t^{\Delta}-t^{\nabla} = -2t$ for the one-orbital model, and $-t-(-t)=0$ for the inversion even/odd interorbital model.
     
     This process to find flat band states of single-orbital model for more than two sublattices necessitates the linear combination of degenerate eigenvectors, leading to an inevitable band touching between the resulting flat band and other dispersive bands, as mandated by symmetry \cite{hwang_flat_2021}. Similarly, the flat bands in the pyrochlore structure also touch other dispersive bands at $\Gamma$ point \cite{wakefield_three-dimensional_2023}. The wavefunction can also be constructed using the two-sublattice eigenvectors ($E_{\mathbb T_2}= -t$) with $w^{\mathbb T_2}_p=1, w^{\mathbb T_2}_q=-1, w^{\mathbb T_2}_r=w^{\mathbb T_2}_s=0$ in the three-fold degenerate irrep ${\mathbb T}_2$ for a tetrahedron belonging to the ${T}_d$ group. Combined with the site symmetry ${D}_{3d}$, there are two sets of two-sublattice solutions out of the three degenerate eigenstates with $E_{1/2/3}^P=-t$: $\vert\Psi^P_1\rangle=\left[1,1,-1,-1\right]/2$, $\vert\Psi^P_2\rangle=\left[1,-1,0,0\right]/\sqrt{2}$, and $\vert\Psi^P_3\rangle=\left[0,0,1,-1\right]/\sqrt{2}$. They form the two degenerate flat bands solution belonging to the site symmetry irrep ${\mathbb E}_u$ with energy $E^P_{FB} = E_{\mathbb T_2}^\Delta+E_{\mathbb T_2}^\nabla = -t^{\Delta}-t^{\nabla} = -2t$ for the one-orbital model, and $-t-(-t)=0$ for the inversion even/odd interorbital model.
     
    \subsection{MEM for Two Orbital Model}
    In the one-orbital model, the hopping term $tf(\mathbf{k})_{ij}c^\dagger_i(\mathbf{k})c_j(\mathbf{k}) = (tf(\mathbf{k})_{ji}c^\dagger_j(\mathbf{k})c_i(\mathbf{k}))^\dagger$.  Thus, there is no extra degrees of freedom to further divide $H^\Delta$ and $H^\nabla$. By contrast, in the two-orbital model, because of the possible orbital combinations $H^{\Delta/\nabla}$ can be further divided into left and right handed chiral hopping channels. For example, the right-handed interorbital chiral hoppings are defined according to $c^\dagger_{O_1,j}(\mathbf{k})c_{O_2,i}((\mathbf{k}))$ with $\{ij\} = \{12\},\{23\},\{31\}$. As a result, the mirror even/odd interorbital Hamiltonian is decomposed as
    \begin{equation}
    H^{oe}_M = H_R^{\Delta}+H_L^{\Delta}+H_R^{\nabla}+H_L^{\nabla},
    \end{equation}
    where 
    \begin{equation}
    H_{R}^{\Delta/\nabla} = t_{R}^{\Delta/\nabla}\left[\begin{array}{cccccc}
    0 & 0 & 0 & 0 & e^{\mp ik_3} & 0\\
    0 & 0 & 0 & 0 & 0 &  e^{\mp ik_1}\\
    0 & 0 & 0 & e^{\mp ik_2} & 0 & 0 \\
    0 & 0 &  e^{\pm ik_2} & 0 & 0 & 0\\
    e^{\pm ik_3} & 0 &  0 & 0 & 0 & 0\\
    0 & e^{\pm ik_1} & 0 & 0 & 0 & 0
    \end{array}\right]
    \label{rc-upH}
    \end{equation}
    and
    \begin{equation}
    H_{L}^{\Delta/\nabla} = t_{L}^{\Delta/\nabla}\left[\begin{array}{cccccc}
    0 & 0 & 0 & 0 & 0 &  e^{\pm ik_2}\\
    0 & 0 & 0 & e^{\pm ik_3} & 0 & 0\\
    0 & 0 & 0 & 0 & e^{\pm ik_1} & 0 \\
    0 & e^{\mp ik_3} &  0 & 0 & 0 & 0\\
    0 & 0 &  e^{\mp ik_1} & 0 & 0 & 0\\
    e^{\mp ik_2} & 0 & 0 & 0 & 0 & 0
    \end{array}\right]
    \label{lc-upH}
    \end{equation}
    Here, $t_R^{\Delta}=-t_L^{\nabla}=t_A$ and $t_R^{\nabla}=-t_L^{\Delta}=t_B$ are required to construct the antisymmetric interorbital Hamiltonian $H_{12}$. The eigenstates for these chiral hopping Hamiltonians around the up and down triangles are the interorbital bonding/antibonding states on the three bonds. They are therefore triple-degenerate two-sublattice wavefunctions. We construct the antisymmetric relation with respect to $\sigma_v$ by combining the mirror related up and down chiral Hamiltonians: $H_A = H_R^{\Delta}+H_L^{\nabla}$ and $H_B = H_L^{\Delta}+H_R^{\nabla}$. If $t_A = t_B$, antisymmetric relation with respect to $\sigma_v'$ also exists because $t_R^{\Delta}=-t_L^{\Delta}$, forming the Hamiltonian $H^{oe}_M = H_A+H_B$. If $t_A = -t_B$, the hopping structure is symmetric with respect to $\sigma_v'$ but antisymmetric with respect to site inversion center, forming the Hamiltonian $H^{oe}_I = H_A-H_B$. As a result, the constructed mutual eigenstates with $\mathbf{k}$-independent eigenvalues for $H_A$ and $H_B$ are: $\vert\Psi^{1}_{A/B}\rangle=\left[e^{\pm ik_1}, e^{\pm ik_2}, e^{\pm ik_3}, e^{\mp ik_1}, e^{\mp ik_2}, e^{\mp ik_3}\right]/N_1$, $\vert\Psi^{2}_{A/B}\rangle=\left[-e^{\pm ik_1}, -e^{\pm ik_2}, -e^{\pm ik_3}, e^{\mp ik_1}, e^{\mp ik_2}, e^{\mp ik_3}\right]/N_1$. These states form flat bands at zero energy, which is a result of the hopping with opposite sign on the up and down triangles. $\vert\Psi^{i}_{A/B}\rangle$ can be linearly combined into orbital pure wavefunctions, which will be the eigenstates when the chemical potential $\mu_{O_1}\neq\mu_{O_2}$.
    
    Next to find the total eigenstate for $H^{oe}_M$ or $H^{oe}_I$. For $O_1$ as an example, we calculate the cross terms $H_A\vert\Psi_B\rangle$ and $H_B\vert\Psi_A\rangle$ and obtain $H_A\vert\Psi_B\rangle = it\left[\mathbf{s}(k_2-k_3), \mathbf{s}(k_3-k_1), \mathbf{s}(k_1-k_2), 0, 0, 0\right]/N$, which is purely imaginary, and  $H_A\vert\Psi_B\rangle =(H_B\vert\Psi_A\rangle)^*$. As a consequence, $H_A\vert\Psi_B\rangle +H_B\vert\Psi_A\rangle = {\rm Real}(H_A\vert\Psi_B\rangle)=0$. Therefore, for the total eigenstate to remain a flat band solution, the parity of the wavefunction should match with the total Hamiltonian:
    \begin{align}
    &(H_A\pm H_B)(\vert\Psi_A\rangle\pm\vert\Psi_B\rangle)\nonumber\\ 
    &= \mu(\vert\Psi_A\rangle+\vert\Psi_B\rangle)\pm(H_A \vert\Psi_B\rangle+H_B\vert\Psi_A\rangle) \nonumber\\ 
    &= \mu(\vert\Psi_A\rangle+\vert\Psi_B\rangle)
    \label{memproof}
    \end{align}
    \subsection{Definition of Up and Down Tetrahedrons}
    \begin{figure}[h]
	\centering
	\includegraphics[width=0.3\columnwidth]{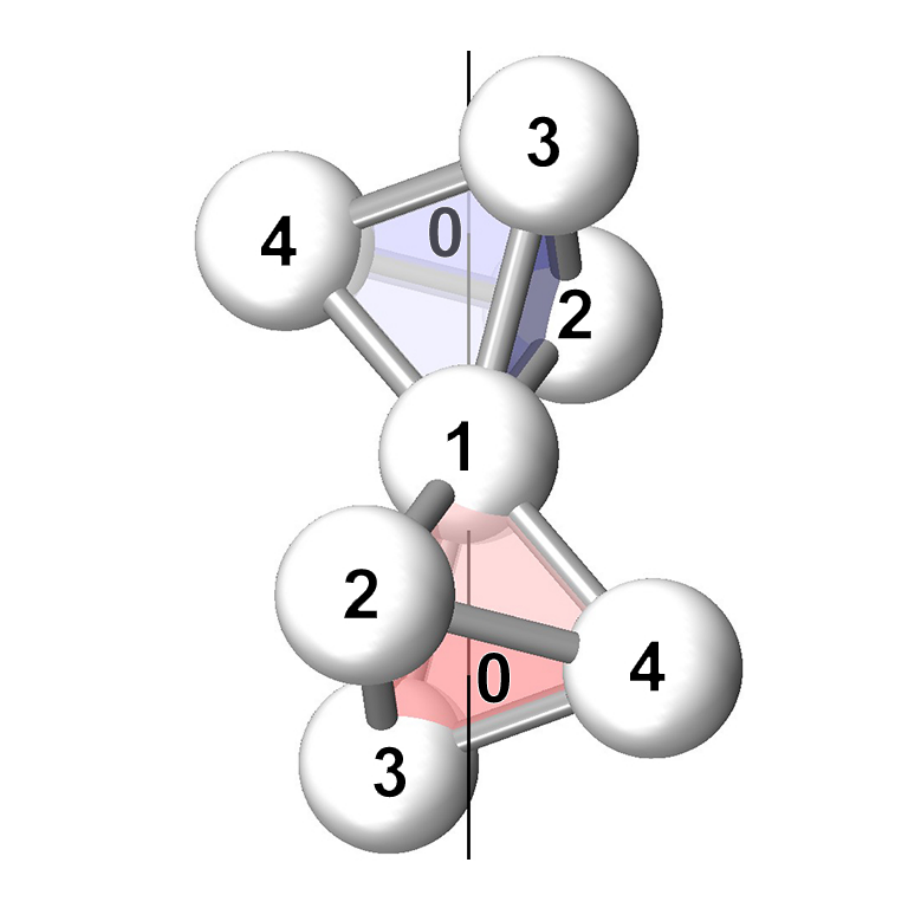}
	\caption{ Up (red) and down (blue) tetrahedrons in pyrochlore lattice. The two tetrahedrons are defined by the different chirality with respect to the vector from the shared vertex to the center of the triangle face consist of the rest of the vertices.}
	\label{figS2}
    \end{figure}
     The tetrahedron is defined in the following way: form a perpendicular line $\mathbf{r_{10}}$ from site-1 to the center of triangle of site-2/3/4. Up tetrahedron is a right-handed triangle 234 around the axis $\mathbf{r_{10}}$, and down tetrahedron is a left-handed triangle 234 around the axis $\mathbf{r_{10}}$.

\end{document}